# Photoemission Spectroscopy and Atomic Force Microscopy Investigation of Vapor Phase Co-Deposited Silver/Poly(3-hexylthiophene) Composites

*L. Scudiero,[a),c)], Haoyan Wei,[b),c)] Hergen Eilers [b),d)]*

[a)] Chemistry Department and Materials Science Program, Washington State University, Pullman, WA 99163, USA

[b)] Applied Sciences Laboratory, Institute for Shock Physics, Washington State University, Spokane, WA 99210, USA

Email Contact: Haoyan Wei, wtaurus@msn.com

# ABSTRACT

Nanocomposite matrices of silver/poly(3-hexylthiophene) (P3HT) were prepared in ultra-high vacuum through vapor-phase co-deposition. Change in microstructure, chemical nature and electronic properties with increasing filler (Ag) content were investigated using *in-situ* XPS and UPS, and ambient AFM. At least two chemical binding states occur between Ag nanoparticles and sulfur in P3HT at the immediate contact layer but no evidence of interaction between Ag and carbon (in P3HT) was found. AFM images reveal a change in Ag nanoparticles size with concentration which modifies the microstructure and the average roughness of the surface. Under co-deposition, P3HT largely retains its conjugated structures, which is evidenced by the similar XPS and UPS spectra to those of P3HT films deposited on other substrates. We demonstrate here that the magnitude of the barrier height for hole injection ($\phi_h^+$) and the position of the highest occupied band edge (HOB) with respect to the Fermi level of Ag can be controlled and changed by adjusting the metal (Ag) content in the composite. Furthermore, UPS reveals distinct features related to the C 2p ($\sigma$-states) in the 5-12 eV regions, indicating the presence of ordered P3HT which is different from solution processed films.

# KEYWORDS

## I. INTRODUCTION

The nanoengineering of hybrid polymer-metal, [1-3] -metal oxide, [4] -semiconductors,[5,6] -polymers,[7,8] and –fullerenes,[9] thin films is a fast developing field of nanotechnology. These new composite materials show promise for a variety of applications in novel organic optoelectronics such as solid state electronics, [10,11] electroluminescent devices and photovoltaics, [12-14] and photodetectors. [15] In particular, poly (3-hexylthiophene) (P3HT), owing to its high drift mobility [16] (up to 0.1 $cm^2 V^{-1} s^{-1}$),[10] has been widely adopted as a hole transporting agent in developing plastic electronics.[10,12] The functionality and performance of these devices are largely dependent on the charge transfer process across the interfacial structure between organic semiconductors (polymer) and filler materials. The offsets of the energy levels of the two materials result in the formation of a heterojunction with the polymer being the electron donor and the filler the electron acceptor. The charge transfer process is a function of the relative energy levels of the HOMO and LUMO derived molecular orbitals in the valence band and conduction band regions, respectively. Studies of electronic structures using ultra-violet photoelectron spectroscopy (UPS) and X-ray photoelectron spectroscopy (XPS) have primarily focused on thin films in the layered configuration. However, the electronic properties of nanoparticle-loaded polymers could be very different from those of the extended interfaces due to distinct structural morphology. Most of polymeric devices are solution processed in order to incorporate nanocrystal fillers into the matrix. However, one of the limiting factors for device performance using this process is inefficient charge transport. The presence of solvent adsorbed on the surface of the nanoparticles impedes the interaction between polymer and nanoparticles and the transfer of charges between the two materials and the transport of electron from nanocrystal to nanocrystal.[17, 18] Vapor phase co-deposition is an alternative route to make metal/polymer blend. This process of producing nanocomposites is solvent free and has been used by Grytsenko,[3] Takele[2] and others.

In the present work, we demonstrate for the first time that P3HT and silver can be simultaneously thermally evaporated to form nanocomposite materials. The co-deposition of metal (Ag) and P3HT from two independent sources was achieved to produce polymeric films with different metal contents. The composite materials were investigated by photoemission spectroscopy (XPS and UPS) and atomic force microscopy (AFM). The surface sensitive analysis techniques are utilized to investigate the chemical nature and electronic states of the newly formed composite. XPS probes the core level binding energies allowing us to determine the chemical reactions at the vast interfaces between Ag and P3HT. UPS characterization provides insightful interfacial electronic information of the metal/polymer composite matrices. Our findings are compared with the layered structures (P3HT/Ag). Finally, AFM images provide morphological information such as surface roughness and Ag nanoparticle size as a function of Ag loading.

## II. EXPERIMENTAL

Ag/P3HT composites for UPS/XPS measurements were formed by evaporating P3HT and Ag simultaneously *in-situ* in a high-vacuum deposition chamber (base pressure of $10^{-8}$ torr) by vapor-phase co-deposition using an electron-beam heated effusion cell (Mantis Deposition Ltd.). The details of the experimental setup are described elsewhere.[19]

The quartz crystal microbalance (QCM) was utilized to estimate the deposition rate and film thickness. Its calibration on P3HT was performed on a layer-structured sample (P3HT on top of Ag substrate) with the attenuation of the intensity of Ag 3d $_{5/2}$ peaks by using the ratio of the sample signal to that of a pure Ag substrate signal.[20] The P3HT deposition was kept at a constant low deposition rate (*ca.* 1Å/min) and the silver deposition rate was varied to alter the metal loadings. Three nominal Ag/P3HT volume ratios were selected including 1:3, 1:1 and 3:1. Since the sticking coefficient of Ag to P3HT is unknown, no attempt was made to correlate the deposition rate ratio to the actual Ag concentration in the sample. The thickness for all samples is estimated to be around 50 nm assuring that photoelectrons only come from the interested composite instead of the underlying substrates. This thickness was also confirmed by the absence of XPS signal from the underlying Si surface when used as substrate.

The deposition chamber is interfaced with the UPS/XPS analysis chamber, ensuring sample deposition, transfer, and measurements all under ultrahigh vacuum without exposure to air. Initially, Ag and P3HT were kept below their melting points. Subsequently they were slowly heated to a temperature above their melting points and kept there for about 1 hr to remove potential impurities and other alien volatiles. A polycrystalline silver foil obtained from Alfa Aesar was cut into squares of about 10x10 mm$^2$ and used as substrates for XPS/UPS characterization. The foils were mechanically polished with a typical metal polish paste (Simichrome Polish), successively cleaned under sonication for several minutes in acetone and isopropanol alcohol, and finally rinsed with isopropanol alcohol and dried in air. The substrates were then immediately loaded into the XPS/UPS system and underwent Ar ion sputtering cleaning for at least 15 min.  A piece of Si substrate was put side-by-side with the silver foil substrate for atomic force microscope (AFM) characterization.

XPS and UPS measurements were performed on a Kratos AXIS-165 multi-technique electron spectrometer system with a base pressure of $5 \times 10^{-10}$ torr.  Achromatic X-ray radiation of 1253.6 eV (MgK$\alpha$) was utilized as the XPS excitation source for acquiring all XPS photoelectron spectra.  The binding energies were calibrated against the Au 4f $_{7/2}$ peak taken to be located at 84.19 eV and Ag 3d $_{5/2}$ peak at 368.46 eV.  Deconvolution of S 2p peaks was performed with CasaXPS, commercial software obtained from Casa software, Ltd.  A precise curve fitting was carried out using Gaussian/Lorentzian curves shape (GL(30)) with a full width half maximum (FWHM) of about 1.4 eV to determine the different chemical species.

UPS data were collected with a homemade He lamp source which produces a resonance line He I (21.21 eV) by cold cathode capillary discharge. The spectra were acquired using a hybrid lens that focused the ejected electrons into the Kratos spectrometer.  A bias of -20V was applied to the sample to shift the spectra out of the nonlinear region of the analyzer (0~10 eV kinetic energy).  The energy resolution was determined at the Fermi edge of the Ag foil to be better than 150 meV.

The AFM images were obtained with a PicoPlus microscope from Agilent Inc. and analyzed using a scanning probe imaging processor (SPIP), commercial software from ImageMetrology A/S.  The images were only processed with planefit unless indicated otherwise in the figure caption.

**III. RESULTS AND DISCUSSION**

The structural morphology in nanocomposite materials has been shown to play an important role in the functionality and performance of devices.[5, 6, 7] Atomic force micrographs (AFM) for two Ag/P3HT composite matrices on Si coupons are illustrated in Figure 1. Silver nanoparticles with irregular shapes embedded in the P3HT matrix was observed for both composite films. The formation of silver nanoparticles can be explained by the strong dissimilarity of the two constituents. Metals usually have low solubility in polymers and they have a strong tendency to aggregate into particles and/or clusters due to their mutual affinity. Metal in polymers is often present in a relatively broad size distribution due to the steric hindrance of polymers. The measured diameter of the visible Ag nanoparticles appears to change from about 50 nm to 100 nm as the Ag/P3HT ratio varies from 1:3 to 3:1 (volume ratio). However, there are many very small nanoparticles or metal clusters present in these composites. The composite surface appears very smooth with height variation less than 1 nm although the size of some of the embedded nanoparticles is quite large. The values of the arithmetic roughness, Ra, for the volume ratios (3:1 and 1:3) are 0.32 and 0.21 nm, respectively as estimated with SPIP. Considering that the film thickness is only about 50 nm, the relatively larger nanoparticle size indicates that the nanoparticles may appear elongated laterally as a result of coalescence of neighboring particles during growth. For the samples with a high Ag loading the AFM image reveals more connecting particles (red dashed lines in Figure 1a) embedded in polymer. This microstructure creates pathways to promote efficient transport of electron from Ag nanoparticle to adjacent Ag nanoparticles. In contrast, samples with lower loading of Ag (Figure 1b) produced much smaller Ag nanoparticles which are less likely to be in direct contact with each other. In the latter system the transport of electrons between Ag nanoparticles is reduced by the physical gap between particles which is filled with polymer.

Figure 2 displays photoemission spectra of Ag 3d, C 1s and S 2p obtained for different Ag/P3HT composites. Figure 2a (top left) shows spectra for the clean Ag foil and for three Ag to polymer ratios (3:1, 1:1 and 1:3). A small static charging was observed on composites with relatively lower metal contents (1:1 and 1:3), which is consistent with their isolated particles nature of Ag as confirmed by AFM. The position of the Ag peak for these two composites was adjusted to correct for a small static charging. The same energy shift was used to correct the C 1s and S 2p spectra (Figure 2b and 2c, respectively). Both C 1s and S 2p spectra exhibit an energy shift of about -0.3 eV when compared to a thin film of P3HT (~ 3.1 nm) vapor deposited on polycrystalline Ag foil.[20] The C 1s spectra of Figure 2b display a single symmetric peak centered at BE = 285.2 eV ± 0.1 eV with no shake-up peak at higher BE of the main line. Although the contribution of carbon involves both conjugated carbon and hexyl side chains, their contributions cannot be resolved in our XPS measurements. A similar energy shift for C 1s was also observed in In/polypyrrole system and was attributed to band bending.[21] Similar to individual deposition of P3HT, the absence of shake-up peaks on the high BE side of the main C 1s line indicates that P3HT largely retain its conjugated nature even during the radical co-deposition with hot metal atoms.[22]

On the other hand the S 2p signal in Figure 2c can be deconvoluted with three distinct peaks. These peaks are seen for the three co-deposited matrices on Ag foil and are assigned to 1) the pristine S in P3HT (164.4 eV), 2) S interacting with Ag nanoparticles (162.7 - 162.2 eV) to form most likely an Ag-S species. In their study of Ag-S interaction Xu et al.[23] found that by depositing Ag atoms on thick sulfur layers supported on Mo(110) at 100K, silver sulfide formed. They labeled it AgS$_y$ with S 2p peak at BE = 163 eV. And 3) a different species resulting from P3HT-Ag interaction (161.7 - 161.2 eV), that could be assigned to Ag$_2$S. This assignment agrees with the work of White et al.[24]. They studied the electrochemical growth of Ag$_2$S on Ag(111) by Coulometric analysis and XPS. The sulfur peak was measured at BE = 161.5 eV. The existence of these three peaks can be explained by the presence of more than one chemical bonding state between Ag and S such as AgS and Ag$_2$S compounds.[25,26] The peaks labeled I and II in figures 2 and 3 were also observed and measured at similar BEs on the layered structure (P3HT on Ag) for film thickness between 0.25 to 1.0 nm.[20] The latter weak feature (peak III) is not very clear in layered structures[20] which may be due to the smaller interfacial area or the absence of mixing of the two vapor species that usually promote stronger chemical interaction. As the P3HT polymer concentration increases from a ratio of 3:1 to 1:3 (Ag/P3HT) the intensity of the peak I from intrinsic S atoms (Figure 2c) increased drastically with respect to the other two peaks (II and III). This increase is expected due to the increased content of polymer in the composite. The energy position of peak I remains the same for all three composites. However, the energy position of peaks II and III shifts to slightly higher BE (~0.5 eV) as the ratio of Ag/P3HT reached 1:3. The energy shift measured for peaks II and III is more difficult to explain since the main peak associated with P3HT does not shift with polymer loading. However, at higher silver loadings the Ag nanoparticle size reached 100 nm on average as seen in Figure 1a for the 3:1 concentration. The AFM image reveals that more Ag particles are in contact with each other (dashed lines). As a result the ratio of volume interface with polymer to volume particle ($V_{interface}$/$V_{particle}$) is small and the Ag nanoparticles provide a direct pathway for electrical conductance. The value of this ratio increases with polymer concentration for the 1:1 and 1:3 composites. Therefore one possible explanation for the energy shift is that as the polymer content increases to the ratio of 1:3 and the size of the Ag particles get smaller as shown by AFM, the negative charge on the S atom and the positive charge on the Ag atom become more localized which is in contrast with a more delocalized charge distribution over the larger available Ag area in the case of the 1:1 and 3:1 volume ratios in which the Ag particles are larger. This localization of charge distribution on the Ag-S reduces the negative charge on the sulfur making it more positive. As a result the energy of the species formed by S atoms in direct contact with Ag nanoparticles will shift to higher BE. Composite samples made on Si substrate exhibit a similar energy shift for peaks II and III between ratio of 1:3 and 3:1 (see Figure 3c).

Since Ag and S have strong interaction, the face-on conformation may be more favorable rather than the edge-on conformation because of the steric presence of hexyl side chains on both sides of the thiophene rings. In our previous work on P3HT film deposited on Ag we found that pristine S appeared even at the first deposition layer (0.25nm). This indicates that only about one or two monolayers of P3HT molecules are in direct contact with Ag surfaces. This is different from tris (8-hydroxyquinolinato) gallium (Ga$q_3$) on Ga where intrinsic Ga peak was only observed after 1.2 nm.[27] Therefore, the relative area intensity of perturbed and unperturbed S indicates that large amount of S should be in contact with Ag materials. The surface area from only large size Ag nanoparticles (~100 nm in diameter) is not sufficient to support this fact, so there must be large amount of small nanoparticles and clusters present in addition of large particles. This is also observed in other metal/polymer composites [28] where polymer matrices usually decrease the mobility of metal particles and strongly prevent their aggregation.

Figure 3 displays photoelectron spectra of Ag 3d, C 1s and S 2p acquired on bare Ag and for two metal/polymer composites on a Si substrate. The photoelectron spectra of Ag 3d and C 1s of the two composites are very similar to those measured on Ag substrate. The curve fitting of the photoelectron spectra of S 2p reveals three peaks with peaks II and III shifting to higher BE with increasing content of P3HT. The presence of these three peaks in both Ag and Si substrates confirms that the Ag nanoparticles are responsible for the existence of the two Ag-S species.

Figure 4 depicts the He I UPS spectra of co-deposited Ag/P3HT composites for three different Ag loadings. The pure Ag and the thin film of P3HT (3.1 nm) spectra obtained by thermally depositing P3HT on the polycrystalline Ag foil are also shown for comparison. In Figure 4a the position of the secondary cutoff edge shifts with increasing P3HT content. This vacuum level shift ($\Delta$) for all samples is measured with respect to clean Ag. This shift is generally attributed to the presence of a dipole moment at the interface of metal/polymer which here indicates a strong chemical interaction between Ag nanoparticles and P3HT due to charge transfer between the two materials as a result of co-deposition. Surprisingly, the Ag/P3HT sample of 1:3 ratio exhibits more up-shift ($\Delta = - 0.9$ eV) than for pure P3HT on Ag ($\Delta = - 0.56$ eV). This energy shift decreases to - 0.8 and - 0.5 eV as the content of Ag increases. The decreasing value of $|\Delta|$ is expected with increasing amount of Ag in the matrix when measured with respect to clean Ag since the material becomes more and more metallic. The same trend is also observed for the composite samples made on Si coupons (see Figure 5). This larger secondary edge shift could be due to the presence of the large amount of Ag-S complex. This results in a larger interfacial area between Ag and S in samples with low metal content.

The knowledge of both the vacuum shift (Δ) and the position of the highest occupied band edge (HOB) allows for the construction of a schematic representation of the band diagram as a function of Ag loading.  The band diagram is shown in Figure 6. The work function of the blend films decreases from 3.77 to 3.37 eV while the barrier height for hole injection $(\varepsilon_V^F)$ increases with polymer content. Similar to layered structures, as P3HT content increases the HOMO peak shift to higher binding energy, from 3.5 eV to 4.0 eV (the right graph in Figure 4).  Unfortunately the energy gap for these new kinds of composite materials is unknown which does not able us to position the conduction band minimum (CBm) in our band diagram (Figure 6) and determine with confidence if these blend films are p or n type materials.  In their study of P3HT-PCBM film adsorbed on PEDOT/ITO from solution, Ta-Chang Tien et al.[29] found a value for barrier height for the blend film 0.74 eV higher than for the pure PEDOT/ITO $(\varepsilon_V^F = 0)$.  Our results follow the same trend.

Finally, three distinct features at 11 eV, 9.4 eV and 7.6 eV are observed between 5 and 12 eV in Figure 5c.  These features are not well resolved in our P3HT thin film deposited on Ag.  This region is attributed to the σ states (C 2p) contributed from main backbones and hexyl side chains of P3HT.  The emergence of these features indicates that ordered structures were formed more or less in the vapor-phased co-deposited P3HT in contrast to spin-coated films where this region in the experimental data appears as a broad featureless band.[30] This may be explained by the forgoing speculation of face-on conformation due to the preferential S-Ag interaction at the vast interfacial contact. However, the calculated DOS using the DFT method by Hao and co-workers[30] reveals three distinct peaks at 11.2, 10 and 7.8 eV, respectively. This is in relatively good agreement with our measurements. The co-deposition with Ag appears to reduce to a certain extent the complexity of the polymer structure to promote a more ordered structure.

The co-deposition of Ag with P3HT forms complicated binding states at the vast interface and different structural morphology as a function of composition. The interfacial complex Ag-S may be considered as a third phase in addition to original constituents and may play an important role in the design of devices. The new composite materials cannot be treated as simple physical mixture of different component since there are vast chemical interactions between species as a result of large interfacial areas and inter-diffusion.

## IV. CONCLUSION

Chemical, electronic and structural information were obtained on vapor-phase co-deposited Ag/P3HT nanocomposites with variation of metal content. Ag exists in the form of nanoparticles embedded in P3HT matrix as confirmed by AFM images. Composites with low metal content exhibit decrease in electrical conductivity due to the spatial distribution of the metal phase which changes the $V_{interface}/V_{particle}$ ratio. XPS indicates that there are at least two chemical bonding states between Ag and S in P3HT. These chemical interactions only occur at the exact contact and do not extend into bulk P3HT matrix. UPS measurements show some similarity in the derived region (0 - 5 eV) with the conventional layered structures (P3HT on top). Large secondary cutoff edge shifts ($\Delta$) to higher binding energies were observed with increasing P3HT content. This is mostly due to the vast interfacial areas between Ag/P3HT where the third phase of Ag-S complex resides. The magnitude of HOB edge (barrier height) increases with polymer content which impedes the charge injection from the Fermi energy of Ag into the valence band of the composite materials. Furthermore, we have demonstrated in this work that it is possible to tune the value of the barrier height ($\varepsilon_F^\pi$) from 0.55 to 1.36 eV by simply varying the composition of the blend film (P3HT/Ag ratio ranging from 3:1 to 1:3) and change the electronic property of nanocomposite materials. Finally, three small yet clear peaks were observed in the 5-12 eV region which is related the $\sigma$-states of P3HT backbone indicating the possible ordering of these vacuum thermally evaporated polymers as a result of thermal annealing.

**ACKNOWLEDGMENTS**

This work was supported by ARO Grant W911NF-06-1-0295 and by ONR Grant N00014-03-1-0247.

<div align="center">**Captions**</div>

**Figure 1:** Tapping mode AFM topological micrographs of Ag/P3HT composites on Si substrates with Ag/P3HT volume ratio of 3:1 (a) and 1:3 (b).

**Figure 2:** XPS spectra of co-deposited Ag/P3HT matrix on Ag foil with different Ag loadings (a) Ag 3d, (b) C 1s and (c) S 2p.  The ratios represent the amount of Ag to P3HT Spectra of vapor-phase-deposited P3HT films on Ag were also superimposed for comparison.

**Figure 3:** XPS spectra of co-deposited Ag/P3HT matrix on silicon wafer with different Ag contents (a) Ag 3d, (b) C 1s and (c) S 2p.  The ratios indicate the amount of Ag to P3HT.

**Figure 4:** He I UPS spectra of co-deposited Ag/P3HT matrix on Ag foil with three different Ag contents.  Vacuum shifts (Δ) and the highest occupied band edge are shown.

**Figure 5**:  He I UPS spectra of co-deposited Ag/P3HT matrix on Si coupons with two different Ag contents.  Vacuum shifts (Δ) and highest occupied band edges are shown.

**Figure 6**:  Schematic energy diagram of the interfacial electronic structure of P3HT-Ag nanocomposite as a function of Ag content.

**Figures**

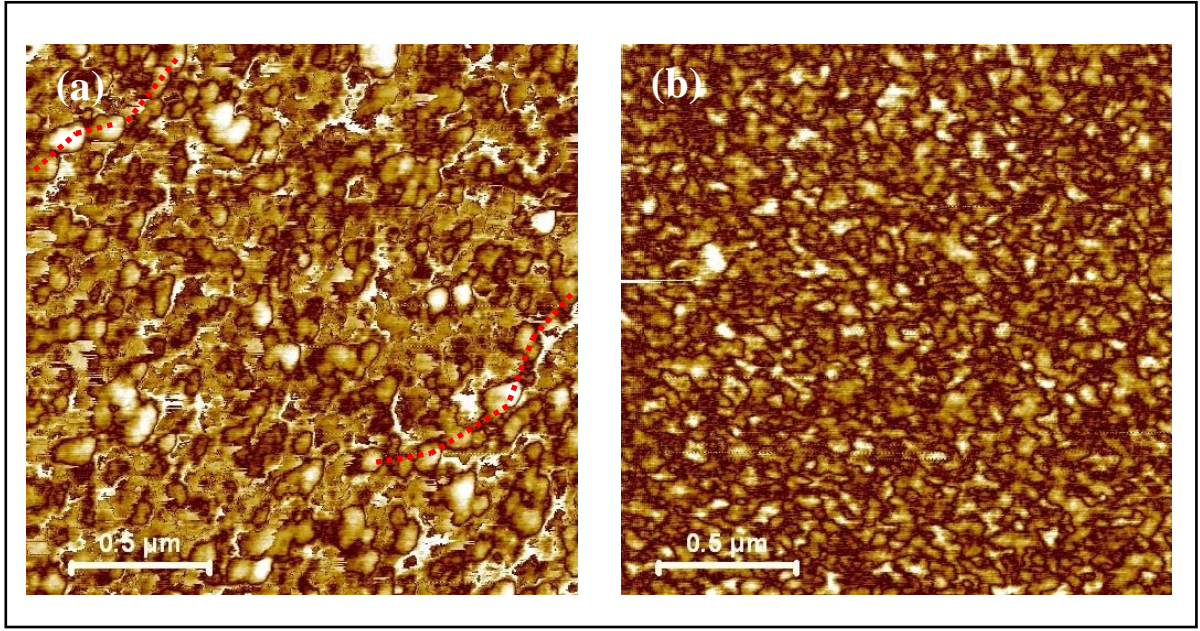

**Figure 1**

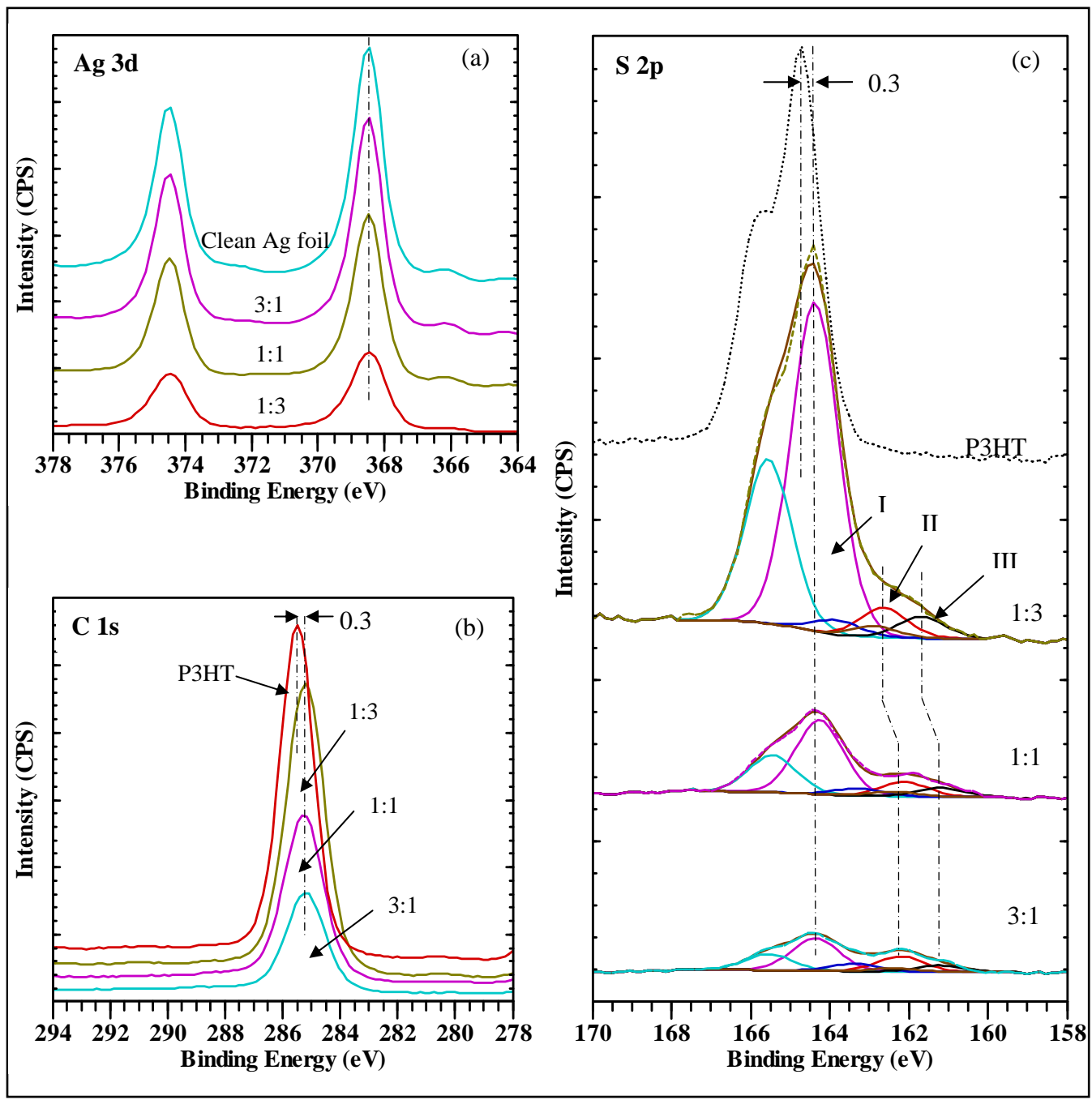

**Figure 2**

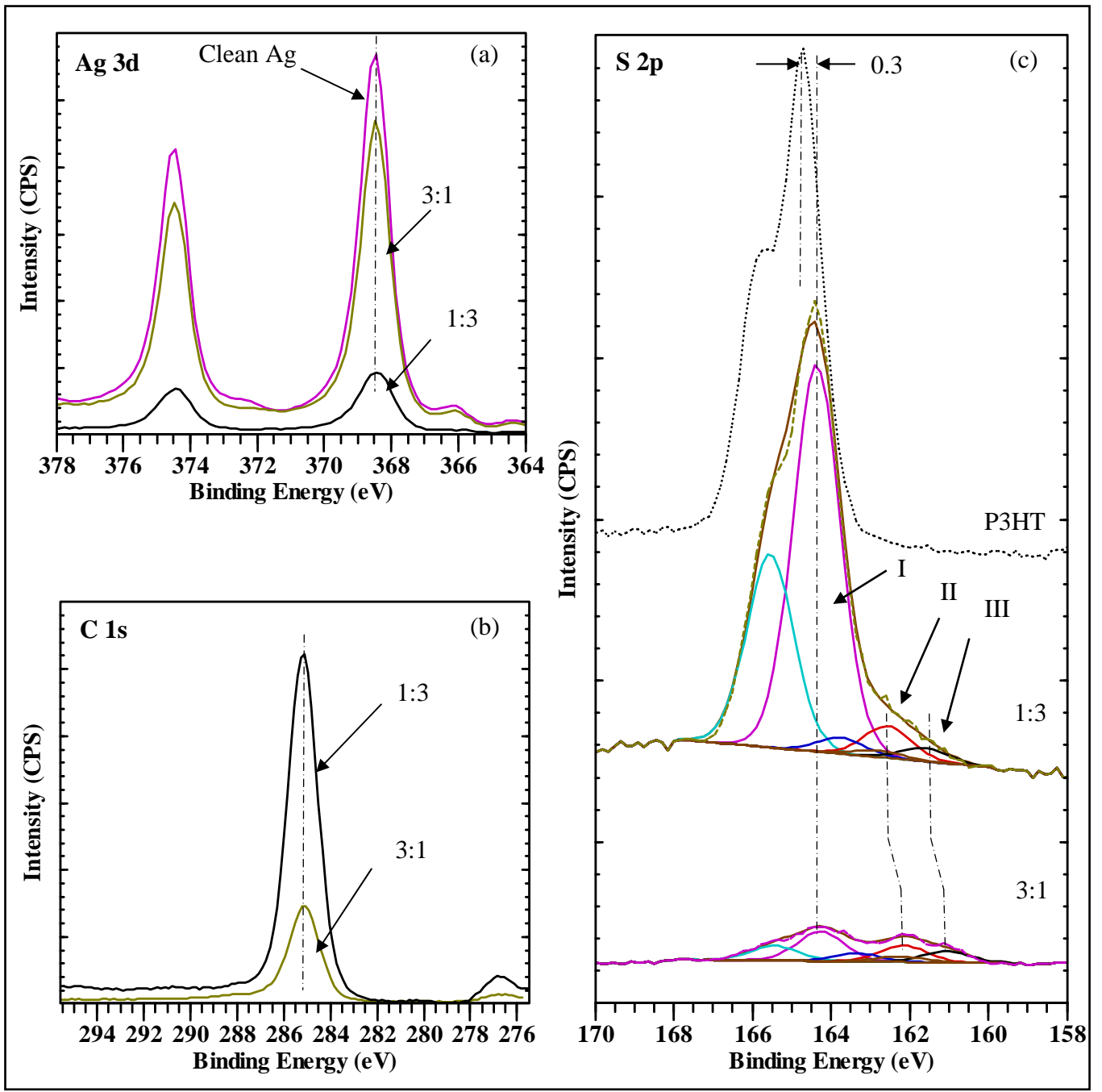

**Figure 3**

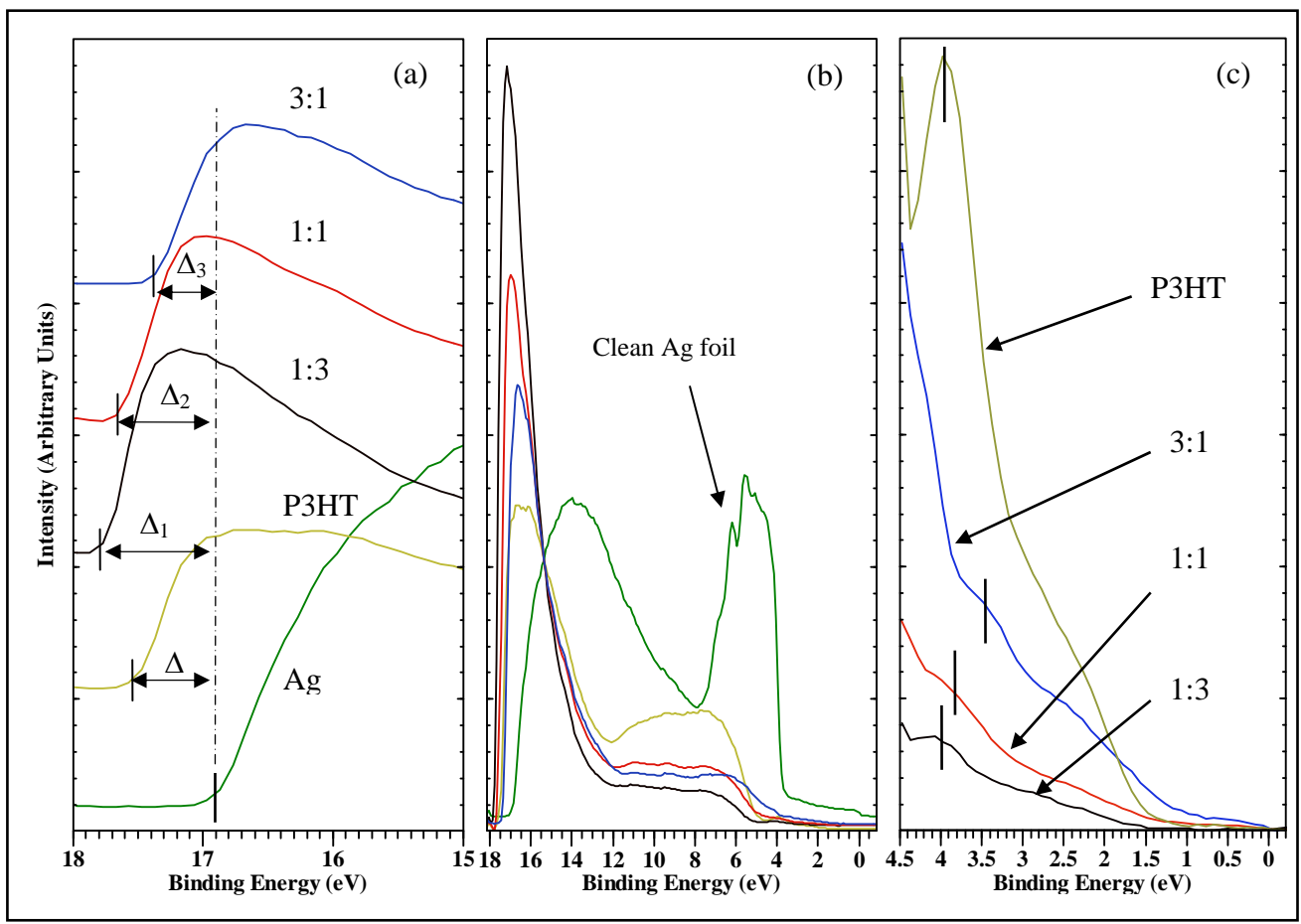

Figure 4

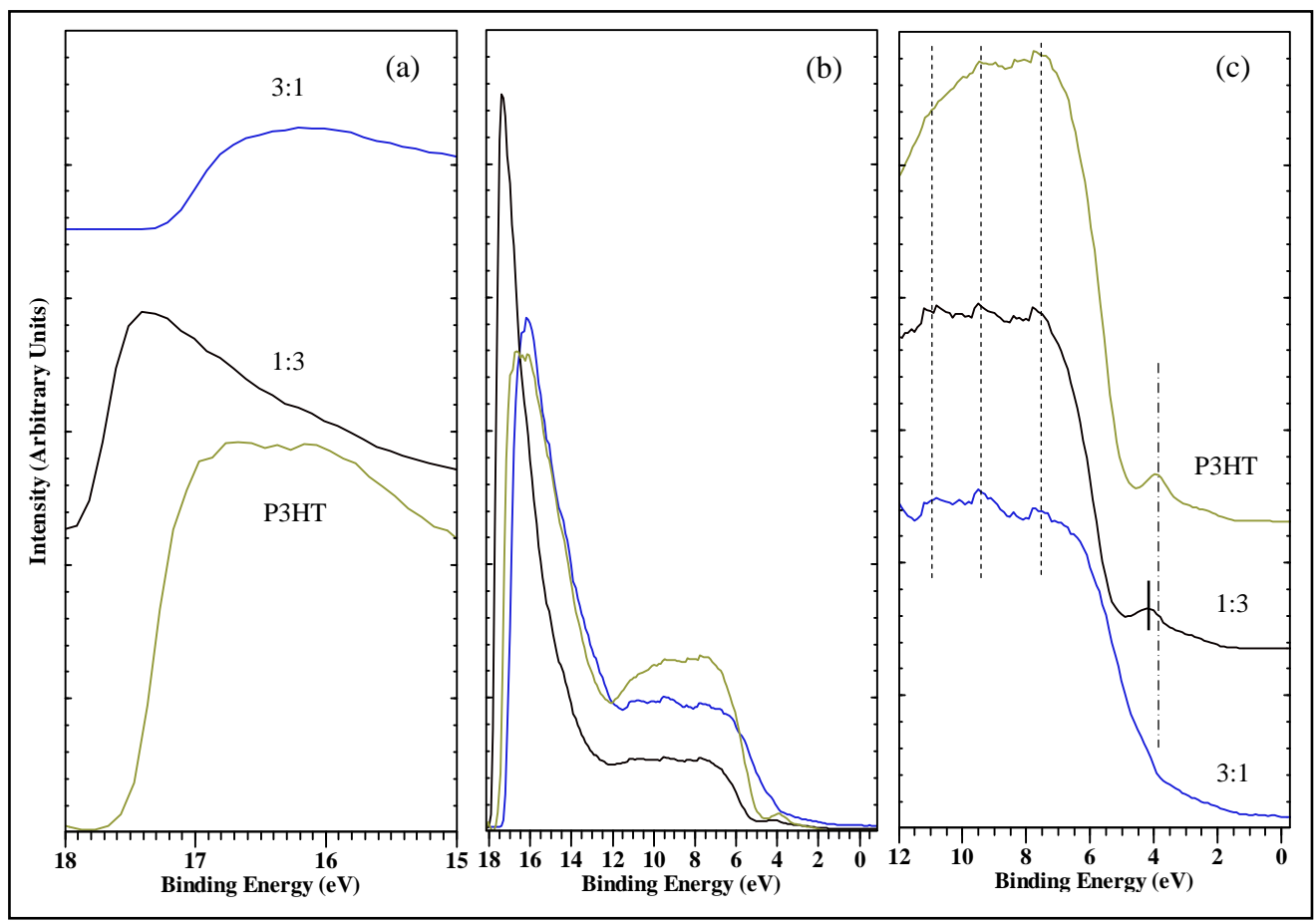

**Figure 5**

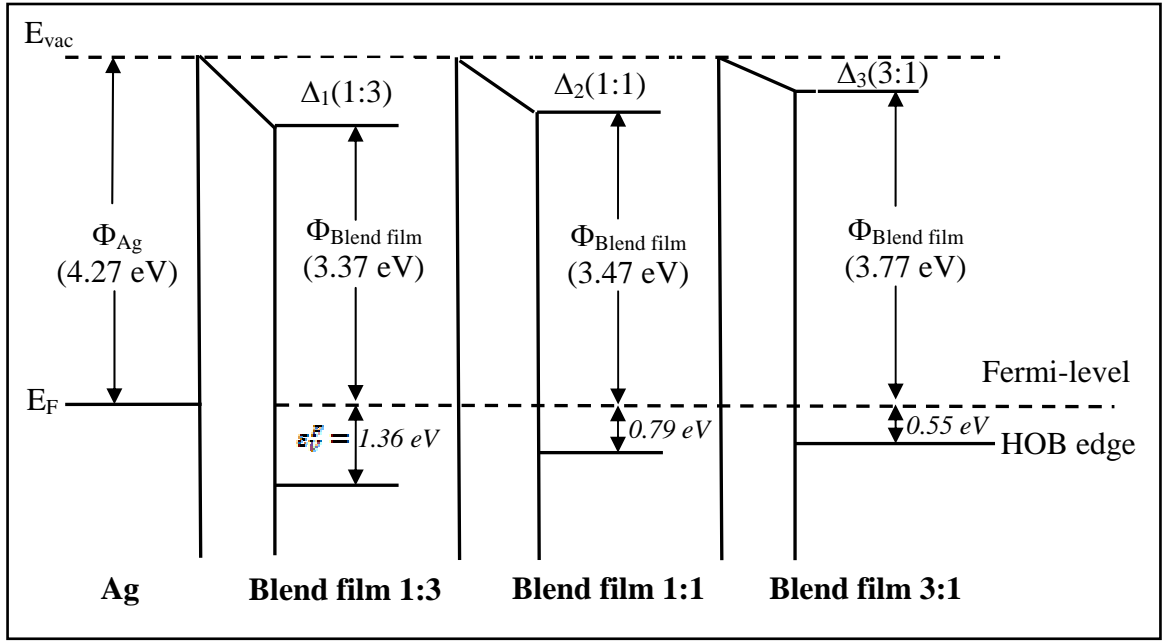

**Figure 6**

# References

[1] A. Biswas, O.C. Aktasa, J. Kanzowa, U. Saeeda, T. Strunskusb, V. Zaporojtchenkoa, F. Faupela, *Mater. Lett.* **58** (2004) 1530– 1534.

[2] H. Takele, S. Jebril, T. Strunskus, V. Zaporojchenko, R. Adelung, F. Faupel, *Appl. Phys. A* **92**, 2008, 345-350.

[3] K. P. Grytsenko, S. Schrader, *Adv. Coll. Interfac. Sci.,* **116**, 2005, 263-276.

[4] Johann Boucle, Punniamoorthy Ravirajan, Jenny Nelson, *J. Mater. Chem.* **17**, 2007, 3141-3153.

[5] N. C. Greenham,* Xiaogang Peng, and A. P. Alivisatos, *Phys. Rev. B* **54**, (1996), 17628-17637.

[6] Wendy U. Huynh, Janke J. Dittner, William C. Libby, Gregory L. Whiting, A. Paul Alivisatos, *Adv. Funct. Mater.* **13**, (2003) 73-79.

[7] A. Biswas, I. S. Bayer, P. C. Karulkar, *J. A. P.* **102**, (2007) 083543.

[8] Marc M. Koetse,a_ Jörgen Sweelssen, Kornel T. Hoekerd, and Herman F. M. Schoo, Sjoerd C. Veenstrab_ and Jan M. Kroon,  Xiaoniu Yang and Joachim Loos, *Appl. Phys. Lett.* **88**, 083504 (2006).

[9] Christoph J. Brabec, James R. Durrant, *MRS Bulletin* **33**, 670 (2008).

[10] H. Sirringhaus, N. Tessler, and R. H. Friend, *Science* **280,** 1741 (1998).

[11] H. E. Katz and Z. Bao, *J. Phys. Chem. B* **104,** 671 (2000).

[12] J. Y. Kim, K. Lee, N. E. Coates, D. Moses, T.-Q. Nguyen, M. Dante, and A. J. Heeger, *Science* **317,** 222 (2007).

[13] G.-M. Ng, E. L. Kietzke, T. Kietzke, L.-W. Tan, P.-K. Liew, and F. Zhu, *Appl. Phys. Lett.* **90**, 103505 (2007).

[14] K. Kim, J. Liu, M. A. G. Namboothiry, and D. L. Carroll, *Appl. Phys. Lett*. **90,** 163511 (2007).

[15] T. Rauch, D. Henseler, P. Schilinsky, C. Waldauf, J. Hauch, and C. J. Brabec, in *IEEE-NANO 2004, Fourth IEEE Conference on Nanotechnology*, Muenchen, Germany, 2004, p. 632.

[16] A. Babel and S. A. Jenekhe, *Macromolecules* **36,** 7759 (2003).

[17] C. Y. Kwong, A. B. Djurisic, P. C. Chui, K. W. Cheng, W. K Chan, *Chem. Phys. Lett.*, 384 (2004) 372-375

[18] Christophe J. Brabec, James R. Durrant, *MRS Bulletin*, 33 (2008) 670-675

[19] A. Biswas, H. Eilers, F. Hidden, O. C. Aktas, and C. V. S. Kiran, *Appl. Phys. Lett.* **88,** 013103 (2006).

[20] H. Wei, L. Scudiero, and H. Eilers, Submitted to *Appl. Surf. Sci.* (2009).

[21] T. Ogama and H. Koezuka, *J. Appl. Phys.* **56,** 1036 (1984).

[22] W. R. Salaneck, O. Inganas, B. Themans, J. O. Nilsson, B. Sjogren, J. E. Osterholm, J. L. Bredas, and S. Svensson, *J. Chem. Phys.* **89,** 4613 (1988).

[23] S. Y. Li, J. A. Rodriguez. J. Hrbek, H. H. Huang, G. Q. Xu, *Surf. Sci.* 395 (1998) 216-228.

[24] Jodie L. Conyers, Jr., Henry S. White, *J. Phys. Chem. B* 1999**,** *103,* 1960-1965.

[25] C. W. Bauschlicher, Jr., H. Partridge, and S. R. Langhoff, *Chem. Phys.* **148,** 57 (1990).

[26] V. K. Kaushik, *J. Electron Spectrosc. Relat. Phenom.* **56,** 273 (1991).

[27] R. Schlaf, C. D. Merritt, L. C. Picciolo, Z. H. Kafafi, *J. Appl. Phys.* **90,** 1903(2001).

[28] H. Wei and H. Eilers, *Thin Solid Films* **517,** 575 (2008).

[29] Chih-Ping Chen, Ta-Chang Tien, Bao-Tsan Ko, Yeu-Ding Chen, Ching Ting, ACS *Appl. Mater. Interf. Lett.* 1, 741 (2009)

[30] X. T. Hao, T. Hosokai, N. Mitsuo, S. Kera, K. K. Okudaira, K. Mase, and N. Ueno, *J. Phys. Chem.* B **111,** 10365 (2007).